# From Classical to Quantum Machine Learning: Different Approaches in Fission Barrier Height Estimation

Serkan Akkoyun[a,b], Cafer Mert Yeşilkanat[c], Paul Stevenson[a]

[a]School of Mathematics and Physics, University of Surrey, Guildford, Surrey GU2 7XH, United Kingdom

[b]Department of Physics, Faulty of Science, Sivas Cumhuriyet University, Sivas, Türkiye

[c]Department of Mathematics and Science Education, Artvin Coruh University, Artvin, Türkiye

**Abstract**

The fission barrier energy is a fundamental property of nuclear structure that governs the stability of nuclei against fission, directly affecting their spontaneous fission half-lives and the formation of superheavy elements. However, because it can only be measured indirectly, it also enables the emergence of alternative, complementary, fast, and accurate prediction tools for traditional theoretical models. In this study, we examine the use of classical, hybrid, and quantum support vector regression (SVR) approaches to estimate fission barrier heights, starting from fundamental nuclear properties and their derived additional properties. For this purpose, eight different SVR-based approaches are considered: (i) Classical SVR, (ii) Enhanced Classical SVR with polynomial and trigonometric feature extensions, (iii) Quantum-Inspired SVR, (iv) Hybrid SVR, (v) Enhanced Hybrid SVR, (vi) Quantum Core SVR, (vii) Fixed-Parameter Quantum Feature Map SVR, and (viii) Pure Quantum SVR. The models were trained and tested on a dataset of 317 isotopes in the Z (atomic number) range of 98–126, encompassing the actinide and superheavy regions. The model performance was evaluated in terms of $R^2$, RMSE, MAE, and training-test $R^2$ gap. The results show that the hybrid model achieves the best overall performance. However, while quantum-enhanced approaches are still limited by circuit depth and optimization precision, they achieve competitive accuracy comparable to classical results. Considering future developments in quantum hardware and algorithms, quantum approaches are expected to achieve considerable improvements. The findings suggest that quantum machine learning can supplement classical approaches and offer a promising path toward more accurate and efficient nuclear property predictions.

1. **INTRODUCTION**

The fission barrier energy is a fundamental nuclear structure quantity that characterizes the stability of a nucleus against fission (Möller & Nilsson, 1970). The barrier height directly influences (i) the channel through which an excited compound nucleus decays between neutron evaporation and fission, (ii) the neutron-induced fission cross-section, (iii) spontaneous fission half-lives, and the probability of superheavy nuclei forming (Möller, et al., 2009; Möller, et al., 2015; Wang, et al. 2019; Kowal, Jachimowicz & Sobiczewski, 2010; Itkis, Oganessian & Zagrebaev, 2002). The double-humped (inner/outer) barrier profile, a characteristic feature of actinides, is a crucial input for the applications of nuclear technology (reactor safety, fuel cycle, criticality) and astrophysical nucleosynthesis scenarios (Bjornholm & Lynn, 1980; Bender, Heenen, & Reinhard, 2003; Schunck & Robledo, 2016). However, this energy is not a measurable quantity and therefore must be determined from indirect experimental results and through calculations based on theoretical models. Therefore, this process is subject to large uncertainties, while precise and computationally accurate fission barrier energy estimation is a fundamental requirement of current nuclear science and technology.

Traditional theoretical approaches developed into two main lines: (i) the macro–micro formalism of the finite-range liquid-drop model (FRLDM) (with Strutinsky shell effects) and (ii) self-consistent mean-field (Skyrme/Gogny HFB) calculations in the energy density functional (EDF) framework (Möller et al., 2016; Bender et al., 2003; Stone & Reinhard, 2007; Dobrowolski, Pomorski & Bartel, 2007; Pomorski & Dudek, 2003; Hofmann, 1974). In both lines, projection of multidimensional potential energy surfaces, identification of the least action path along the action integral, and estimation of amounts such as the inertia tensor demand high computational costs and parameter uncertainties. Accordingly, the predictions of different models can differ significantly, especially in the heavy and superheavy regions.

While quasi-experimental "double hump" barrier parameters (inner/outer barrier heights and widths) for actinides are available in the IAEA's RIPL library (Capote et al., 2009), the data are limited, and for most isotopes, the evaluations are based on assumptions. This increases the need for fast and accurate estimation tools for barrier height determination. In recent years, machine learning (ML) has provided powerful tools for estimating many quantities in nuclear physics, such as nuclear masses, half-lives, decay energies, charge radii, and reaction observables (Gazula, Clark, & Bohr, 1992; Utama, Piekarewicz, & Prosper, 2016; Niu & Liang, 2018; Yüksel, Soydaner, & Bahtiyar, 2024; Akkoyun, 2020; Amrani, Yesilkanat & Akkoyun, 2024; Akkoyun, Yeşilkana & Bayram 2024; Bayram, Yeşilkanat & Akkoyun 2023). Literature reviews indicate that ML can be used as an alternative to theoretical models. Promising studies (Akkoyun, S., & Bayram, T. 2014; Yesilkanat & Akkoyun 2023) have recently been conducted with high prediction accuracy for ML estimation of fission barrier energies for both actinide and superheavy regions using classical ML algorithms.

In the present study, we focus on the fission barrier energy estimation using SVR-based methods using fundamental nuclear parameters of the nucleus and their derived properties. For this purpose, classical, hybrid, and quantum SVR methods were trained and tested on the same data with the same initial parameters. The methods primarily utilize A (atomic mass), N (number of neutrons), and the following derived properties: N−Z, N/Z, asymmetry ((N−Z)/A), fissility (($Z^2$)/A), log$A$, and the interaction term Z·N. This physics-based 8-dimensional feature space is explicitly defined and scaled for ML computations. Then, eight SVR configurations with different approaches are trained: (1) Classical SVR (simple 8 features), (2) Enhanced Classical SVR (polynomial/trigonometric expansion of features), (3) Quantum-inspired SVR (features from quantum circuit), (4) Hybrid SVR (classical+quantum feature), (5) Enhanced Hybrid SVR, (6) Quantum Core SVR, (7) Fixed-Parameter Quantum Feature Map SVR, and (8) Pure Quantum Core circuit (Biamonte, et al., 2017).

In this paper, Section 2 describes the dataset content and the classical, hybrid, and quantum approximation methods based on SVR. Section 3 provides the results of the eight approaches, a comprehensive discussion comparing the results, and evaluation of their performance. Finally, a general overview emphasizes that ML approaches can be an alternative tool, designed not to replace physically-meaningful theoretical models such as FRLDM and EDF/HFB, but rather as a complementary rapid estimation tool that minimizes uncertainty and enables model-guided exploration.

## 2. MATERIAL and METHODS

### 2.1. Dataset for the ML calculations

The dataset (Kowal, Jachimowicz & Sobiczewski, 2010) used in this study contains fission barrier energy values for heavy and superheavy nuclei, including nuclei ranging from Z = 98 (Cf) to Z = 126 (Ubh), and exhibits a broad distribution in the range A≈232−318. This dataset provides a sufficient number of test areas for ML calculations. Fission barrier values range approximately 0–7 MeV, with a general trend toward decreasing barrier height as Z increases. This is expected due to increased Coulomb repulsion. As A increases, the barrier height first increases and then decreases again. This indicates that the barrier decreases through reduced shape stability as more neutrons are added. The decrease in barriers to 2–3 MeV in the superheavy region indicates that these nuclei will have very short fission lifetimes.

The performance of machine learning models depends not only on the fundamental nuclear variables (Z, N, and A), but also on derived features enriched with physical and non-physical information. Machine learning input parameters were diversified by deriving quantities other than Z, N, and A from the dataset. These new quantities are (N-Z), N/Z, (N-Z)/A, $Z^2$/A, Log A, and (Z.N). All attributes were normalized using the *StandardScaler* (μ = 0, σ = 1), which standardizes each feature by removing the mean and scaling to unit variance (Pedregosa et al., 2011). The training-test split was randomly made at 80%-20%, thus ensuring a fair comparison of all SVR models (classical, hybrid, and quantum-based) with the same initial conditions. The correlation matrix in Fig. 1 was computed using the Pearson correlation coefficient. The resulting 9×9 matrix reflects the pairwise linear dependence structure between the selected nuclear features. The Pearson coefficients were obtained using the *corr()* function in Python's *pandas* library, which evaluates linear relationships between variables based on the Eq.1. Here, $x_i$ and $y_i$ denote the individual observations of variables $X$ and $Y$, respectively, while $\bar{x}$ and $\bar{y}$ represent their sample means.

$$r_{xy} = \frac{\sum(x_i-\bar{x})(y_i-\bar{y})}{\sqrt{\sum(x_i-\bar{x})^2(y_i-\bar{y})^2}} \qquad \text{Eq.1}$$

As can be seen, there is an almost perfect positive correlation between some variables as in A and Z*N. Similarly, a very high correlation was observed between the asymmetry and the N/Z parameter, indicating that nuclear asymmetry is directly related to the neutron-proton ratio. On the other hand, strong negative correlations were noted between the fissility and the N/Z and asymmetry parameters. On the other hand, the fission barrier shows a weak negative correlation with several parameters, with the most significant correlation being with the fissility. The matrix is a tool for understanding the statistical relationships between the fundamental variables used in nuclear models and fission processes.

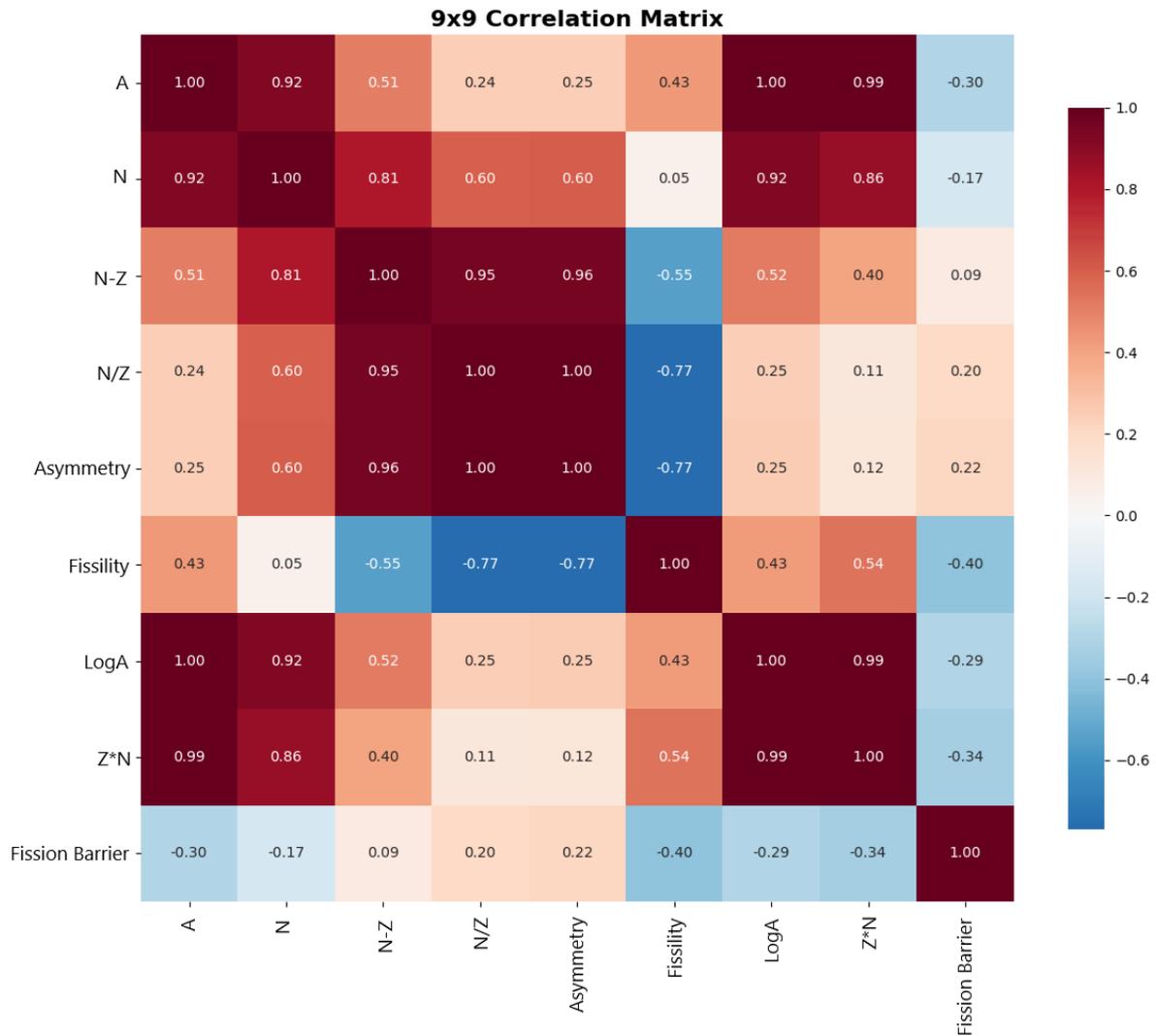

**Fig. 1** Pearson correlation matrix showing the relationships between various nuclear parameters

## 2.2 Different ML Approaches

To estimate the fission barrier energy, three different approaches, classical, hybrid, and quantum, are considered in the SVR model. To ensure fair comparisons between all models, the same initial SVR parameters are used in all models. Following the notation of Pedregosa et al. (2011), these parameters are C: [1, 10, 100, 1000, 5000], gamma: ['scale', 'auto', 0.0001, 0.001, 0.01, 0.1], epsilon: [0.001, 0.01, 0.05, 0.1, 0.15], and kernel: ['rbf', 'poly', 'sigmoid']. In the SVR optimization, a parameter grid search is performed over a wide range of C, gamma, epsilon, and kernel types. C represents the regularization parameter controlling the trade-off between model complexity and training error; gamma denotes the kernel coefficient determining the influence range of individual data points; epsilon specifies the width of the insensitive margin in the SVR loss function; and kernel defines the transformation method applied to map data into a higher-dimensional feature space. Though the initial grid consisted of discrete candidate values, the random search process revealed the best performing hyperparameters.

The kernel function was chosen from standard options such as radial basis function, polynomial, and sigmoid kernels, which are internally calculated during training. These kernels map the input feature space to higher-dimensional representations to capture nonlinear relationships. In addition to the classical SVR approach, a quantum support vector regression (QSVR) model was employed in the study. QSVR leverages a quantum feature map to encode classical input data into a quantum Hilbert space. The quantum circuit generates quantum states whose pairwise inner products define the quantum kernel, representing the similarity between data points in a high-dimensional quantum feature space. This allows the model to capture complex nonlinear structures that may not be easily accessible to classical kernels. In the QSVR framework, the quantum circuit was executed for each data sample pair to compute quantum state overlaps. These similarity values were then collected and stored as a quantum kernel matrix, which replaces the classical kernel in the SVR optimization process. The SVR algorithm used this prepared matrix directly as the kernel. Kernel values were calculated from the inner products of the quantum states prepared for the data samples, providing a quantum-enhanced similarity measure. Despite using the same hyperparameter ranges, SVR models optimized in different feature spaces used different numbers of support vectors. This is inherent to the underlying mechanism of SVR, as the number of support vectors depends on both the chosen hyperparameters and the geometry of the feature space. Flowcharts of the approaches used are briefly shown in Fig. 2.

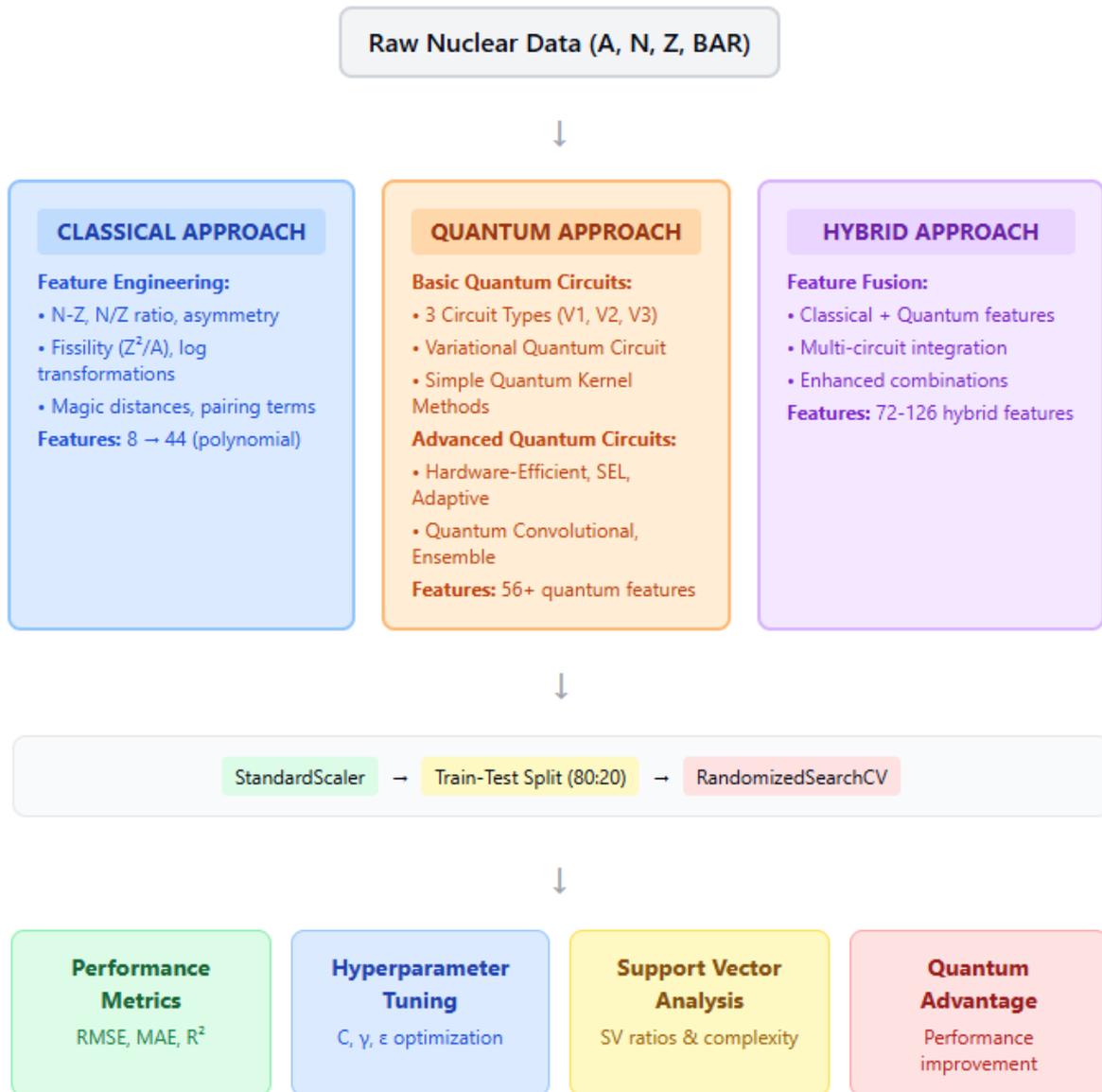

**Fig. 2** Workflow diagram of classical, quantum, and hybrid approaches for nuclear fission barrier estimation.

### 2.2.1 Classical SVR (8 Features Baseline)

Support Vector Regression (SVR) adapts the Support Vector Machine framework to regression problems (Drucker et al., 1997; Smola & Schölkopf, 2004; Cortes & Vapnik, 1995). While SVM finds a maximum-margin hyperplane for classification, SVR generates a function that continuously approximates the target values by preserving the predictions in an ε-insensitive tube, where ε defines a tolerance zone in which prediction errors are not penalized. The ε-insensitive loss function provides robustness to small deviations by treating prediction errors smaller than ε as zero. To capture nonlinear relationships in nuclear data, kernel functions transform the input domain to higher dimensions. This study employs three kernels: the Radial Basis Function (RBF/Gaussian) kernel for smooth nonlinear patterns, polynomial kernels for explicit polynomial relationships, and sigmoid kernels for neural network-like decision boundaries. Hyperparameter optimization was conducted using a random search strategy, systematically sampling combinations of the regularization parameter (C), kernel coefficient ($\gamma$), and tolerance ($\varepsilon$) to identify the configuration that achieved the best predictive performance.

**2.2.2 Enhanced Classical SVR**

The classical SVR model directly utilizes eight physics-based features. However, these features alone may not fully represent nonlinear dependencies in the fission barrier. Nuclear structure effects such as shell closures and pairing correlations can exhibit quasi-periodic oscillatory behaviour across the nuclear chart. To better capture such nonlinear and oscillatory patterns, the feature space was expanded using polynomial features (degree = 2, including squared terms and pairwise products) and trigonometric transformations. With this extended feature representation, the initial eight-dimensional feature space was expanded to 52 features in the Enhanced Classical SVR approach.

**2.2.3 Quantum Inspired + SVR**

The enhanced classical SVR model is based on explicitly designed features via polynomial and trigonometric expansions. As an alternative to this feature enrichment, the Quantum Inspired + SVR method uses quantum-inspired feature extraction, implemented using parameterized quantum circuits (Fig.3). These are simulated in classical hardware via PennyLane's default qubit device (Bergholm et al., 2018; Pedregosa et al., 2011, Ding et al, 2022). The quantum feature extraction process consists of three main steps. First, each of the eight classical features is encoded into qubit rotation angles ($RY(x_i \pi)$, where $x_i$ represents the *i*th normalized feature value) via *RY* gates, and the classical data is mapped to quantum state amplitudes. Second, CNOT gates create entanglement between adjacent qubits, allowing the circuit to capture the relationships between features indirectly through quantum superposition rather than directly in polynomial terms. In the third step, the Pauli operator expectation values $\langle Z \rangle$, $\langle X \rangle$, and $\langle Y \rangle$ are extracted from each qubit, and multiple circuit variants with different gate configurations generate various quantum-derived features, yielding a 56-dimensional feature vector. The quantum-derived features were standardized by subtracting the mean and scaling to unit variance before being used in the SVR model. This ensures that all features have comparable numerical ranges and prevents any single feature from dominating the learning process. The underlying hypothesis of this approach is that entangled quantum states can encode nonlinear feature interactions in ways different from polynomial expansions. However, several important limitations must be acknowledged. This implementation uses classical simulation, not actual quantum hardware. The complexity of simulating quantum circuits scales exponentially with the number of qubits in classical hardware. For the 8-qubit system used here, classical simulation remains applicable but does not offer computational advantages over classical feature engineering. Nevertheless, evaluating quantum kernels in simulation is still valuable, as it allows us to assess whether quantum representations have the potential to outperform classical methods once scalable, fault-tolerant quantum hardware becomes available. In this sense, simulated QSVR serves as a forward-looking feasibility study for identifying problem classes where practical quantum advantage may emerge in the future.

Different quantum circuits were used in the study, as shown in Fig. 3. Circuit-1 is implemented by embedding classical inputs through $RY(\theta i)$ and $RZ(\phi i)$ rotations on four qubits, followed by a linear entanglement chain of CNOTs ($q0 \rightarrow q1, q1 \rightarrow q2$, $q2 \rightarrow q3$). This configuration introduces both amplitude and phase encoding while enabling nearest-neighbour correlations. The final Pauli-Z measurements yield 16 quantum features, capturing single-qubit and multi-qubit interactions up to the four-body level. Circuit-2 is implemented by first embedding classical inputs via $RY(\theta i)$ on each qubit, applying Hadamard gates to create superposition, enacting symmetric ring entanglement through CNOTs ($q0 \rightarrow q1, q1 \rightarrow q2$, $q2 \rightarrow q3, q3 \rightarrow q0$), and finally performing a second $RY(\theta i')$ embedding (data re-uploading). This configuration is chosen to capture both short- and long-range correlations and to enrich the feature space measured via Pauli-Z expectation values. Circuit-3 is implemented using amplitude encoding of classical inputs across four qubits, followed by a set of cross-CNOT entangling gates that couple non-neighbour qubits. Additional layers of single-qubit rotations are applied to redistribute encoded information, and a broader set of Pauli-Z observables is measured. This design yields 24 quantum features, providing a richer representation of higher-order correlations compared to Circuits 1 and 2.

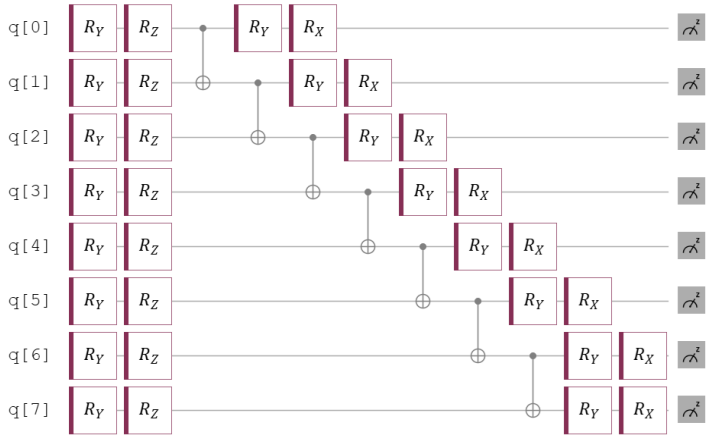
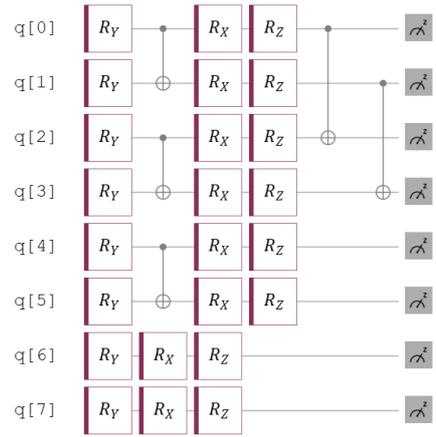
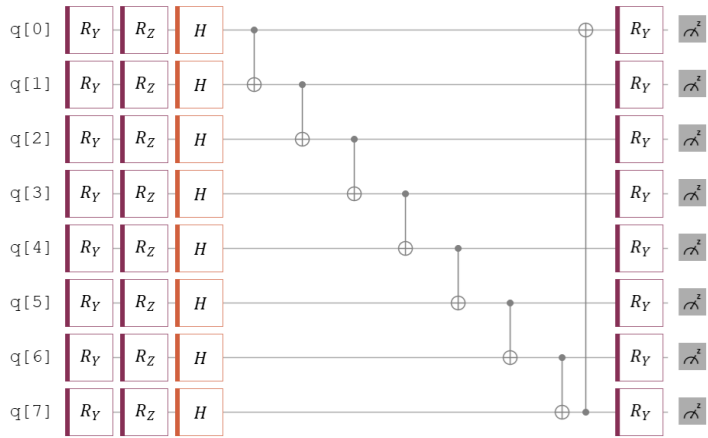

**Fig. 3** Quantum-inspired feature extraction process. Three-parameter quantum circuit variants (1–3) encode eight classical nuclear properties into quantum states using different rotation schemes and entanglement topologies (linear, ring, pairwise, and cross-CNOT).

### 2.2.4 Hybrid SVR (Classical + Quantum Combined)

The hybrid approach combines eight physically interpretable classical features with 56 quantum-derived features to create a 64-dimensional representation. This design aims to preserve the insights gained from nuclear physics while enriching the feature space with quantum-inspired representations (Mitarai, et al., 2018). While classical features maintain direct physical interpretability, quantum features can capture implicit nonlinear correlations. This approach allows us to assess whether quantum features provide complementary information to classical features or merely add unnecessary complexity. A performance comparison between classical-only, quantum-only, and hybrid models reveals the practical contribution of each feature type to prediction accuracy. In the enhanced hybrid model, the total number of features are 108.

**2.2.5 Quantum Kernel SVR (Pre-Computed Kernel)**

The quantum kernel approach constructs a similarity measure by encoding feature pairs into a parameterized quantum circuit and computing state overlap as a kernel function for SVR (Cammarosano Hidalgo, 2025; Havlíček et al., 2019; Schuld & Killoran, 2019; Rebentrost, Mohseni, & Lloyd, 2014; Schölkopf, Smola & Müller, 1998). Kernel values representing similarity between data points are computed as the overlap of quantum states as in Eq. (1).

$$K(x_i - x_j) = |<\psi(x_i)|\psi(x_j)>|^2 \quad (1)$$

where $|\psi(x)\rangle = U_\Phi(x) |0\rangle$. In our implementation (Fig. 4), classical features are encoded via data-dependent $RY$ rotations (dark gates), while $RZ$ gates with randomly initialized but fixed angles (light gates) act as non-trainable phase offsets to increase expressivity. A sparse CNOT block among the top register introduces entanglement, a barrier separates blocks, and a final $RY$ layer completes the map. The panel in Fig. 4 depicts only the feature map $U_\Phi(x)$; kernel values are computed in a separate fidelity circuit implementing $U_\Phi^\dagger(x_i)U_\Phi(x_j)$ with a $|0\ldots0\rangle$ projector (the Z-measurement icons in the figure are illustrative). Measurement-outcome statistics from states generated by different inputs implicitly reflect their overlap: similar distributions yield larger inner products (higher kernel values), while dissimilar ones yield smaller values (Havlíček et al., 2019; Schuld, 2021). For numerical stability we add a small diagonal regularizer $10^{-8}I$ to the kernel matrix. Due to computational limits, the training uses 100 samples. Despite these precautions, the quantum-kernel SVR underperforms strong classical baselines, indicating that for this dataset the fidelity-based similarity does not surpass a standard RBF kernel.

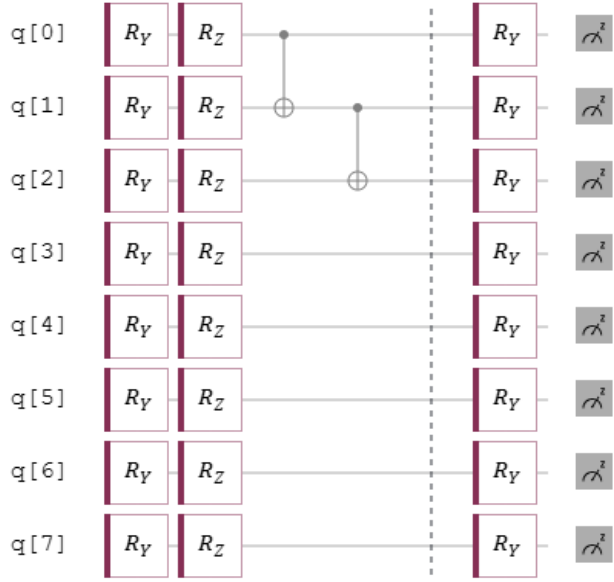

**Fig. 4** Quantum feature map with data-dependent *RY*, fixed *RZ*, and sparse CNOT entanglement. Kernel computed via fidelity circuit.

**2.2.6 Fixed-Parameter Quantum Feature Map + SVR**

Fixed-parameter quantum circuits provide a way to generate nonlinear quantum feature maps without optimizing variational parameters. Rather than updating gate angles during training, we sample the circuit parameters once at initialization and keep them fixed throughout the entire procedure — that is, for all data samples and for all folds in the cross-validation process. This turns the circuit into a static feature embedding rather than a learnable VQC (cf. Mitarai et al., 2018; Benedetti et al., 2019; Peruzzo et al., 2014). In our implementation (Fig. 5), classical features are embedded by a first block of data-dependent RY–RZ–RX rotations, followed by a second block of randomly initialized, non-trainable *RY/RZ* phase offsets. Entanglement is introduced by a linear CNOT chain in the first block and a partially connected pattern in the second; the two blocks constitute a two-layer architecture. After execution, we collect measurement outcomes (single-qubit Z/X/Y expectations and a small set of two-qubit correlators), remove low-variance components, and standardize the resulting quantum features before feeding them to the SVR. While parameterized circuits can learn data-dependent transformations, fixing the parameters prevents adaptation to the underlying data distribution. This setup therefore serves as a fixed-parameter quantum feature-map baseline for fair comparison against the classical, hybrid, and quantum-kernel SVR variants evaluated in this work.

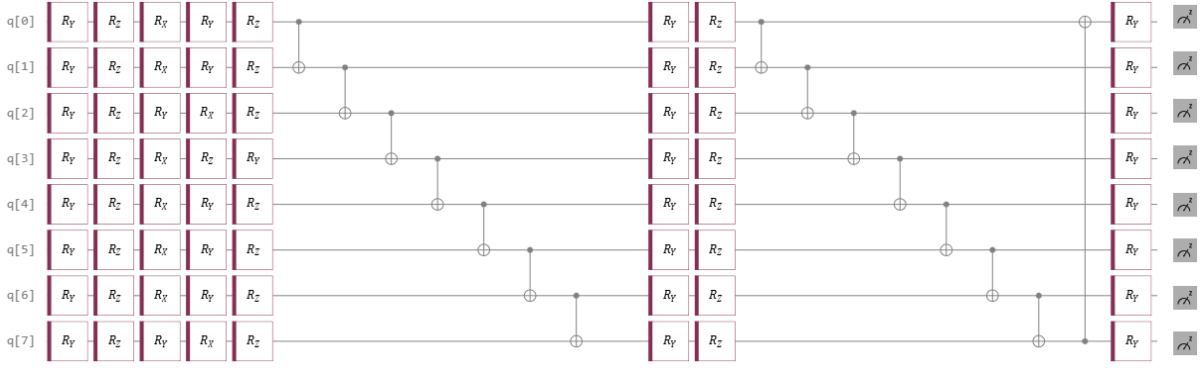

**Fig. 5** Fixed-parameter quantum circuit architecture used to generate quantum feature maps. Circuit parameters are initialized randomly and kept constant throughout SVR training.

### 2.2.7 Pure Quantum Kernel SVR

The quantum kernel approach defines sample similarity through quantum state overlap $K(x,x') = |\langle\psi(x)|\psi(x')\rangle|^2$, where $\psi$ represents the quantum state prepared by parameterized feature maps. This implementation employs an ensemble of four distinct quantum circuits (Fig.6), each utilizing different quantum gate combinations and entanglement strategies: (1) A Hardware-Efficient Ansatz (35%) employs Hadamard initialization, parameterized *RY* and *RZ* rotations with fixed randomly-initialized variational parameters (3×n_qubits), data re-uploading (Pérez-Salinas et al., 2020) via amplitude and phase encoding (*RY*, *RX*), circular CNOT chains, star-topology entanglement from central qubit, and controlled-*RZ* (*CRZ*) gates for pairwise feature interactions (Fig.6). (2) A Strongly-Entangling Circuit (25%) uses Hadamard gates for superposition, $R_Y$ rotations for data encoding ($2\pi$ scaling), all-to-all controlled-RY (CRY) gates for $O(n^2)$ pairwise entanglement, exponentially weighted *RZ* phase encoding, and staggered CNOT layers for additional correlations (Fig.7). (3) An Adaptive Circuit (25%) features conditional initialization (Hadamard or X+Hadamard depending on feature magnitude), norm-dependent circuit depth (2–3 layers), adaptive gate selection (*RY* vs *RX*) depending on data statistics, alternating linear and circular CNOT patterns per layer, and data-norm-scaled *RZ* rotations (Fig.8). (4) A Quantum Convolutional Circuit (15%) implements Hadamard initialization, multi-scale *RY* encoding across 2 two convolutional layers, local 3-qubit kernels with stride-1 convolution via sequential CNOTs, controlled-*RZ* (*CRZ*) operations for local interactions, and controlled-*RY* (*CRY*) pooling on adjacent qubits (Fig.9).

The ensemble kernel is $K_{ensemble} = 0.35·K_1 + 0.25·K_2 + 0.25·K_3 + 0.15·K_4$, combining circuit outputs via weighted average. Here, the terms $K_1$, $K_2$, $K_3$ and $K_4$ are the kernel matrices obtained from different quantum circuit architectures (K1: Hardware-efficient ansatz, K2: Strongly entangling circuit, K3: Adaptive circuit, K4: Quantum convolutional circuit). The weights were determined empirically through preliminary experiments to balance the contribution of each circuit type. Specifically, the hardware-efficient and strongly entangling architectures showed stronger predictive performance in initial validation, leading to their higher weights, while the adaptive and quantum-convolutional circuits were assigned smaller weights. These weights were not optimized during training but were selected prior to final runs to provide a reasonable ensemble balance. The precomputed kernel matrix is fed to SVR (kernel='precomputed'), optimizing only regularization C and tolerance ε. This pure quantum ensemble approach, without classical kernel blending, tests whether sophisticated multi-circuit ensembles can capture nuclear structure correlations, though the computational cost is roughly two orders of magnitude higher than single-circuit implementations due to ensemble evaluation.

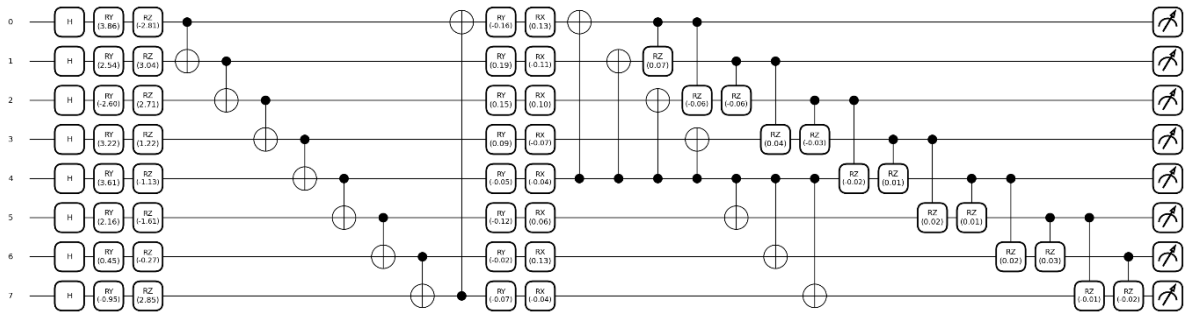

**Fig. 6** Hardware efficient (35%) pure quantum kernel circuit. Each classical data point is encoded into a quantum state *ψ(x)* using gates and CNOT chain.

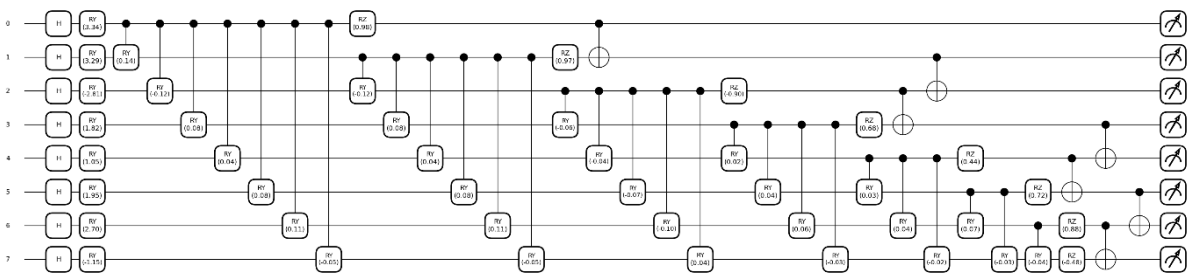

**Fig. 7** Strongly entangled (25%) pure quantum kernel circuit. Each classical data point is encoded into a quantum state *ψ(x)* using gates and CNOT chain.

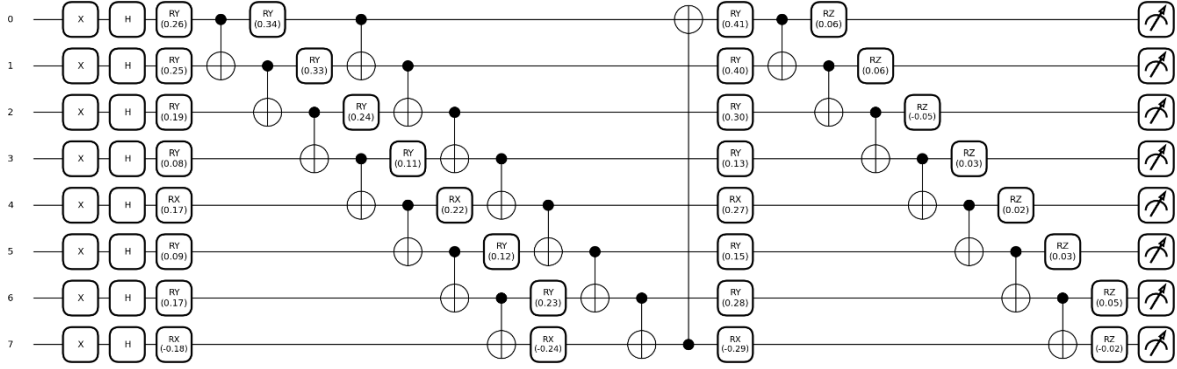

**Fig. 8** Adaptive (25%) pure quantum kernel circuit. Each classical data point is encoded into a quantum state $\psi(x)$ using gates and CNOT chain.

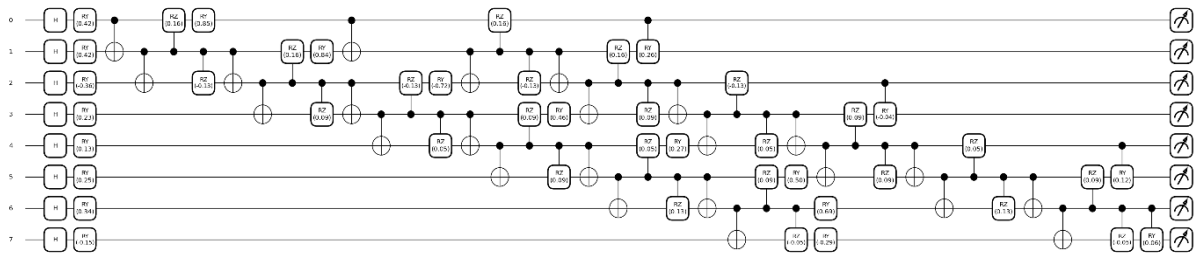

**Fig. 9** Convolutional (15%) pure quantum kernel circuit. Each classical data point is encoded into a quantum state $\psi(x)$ using gates and CNOT chain.

Since one of the main motivations of this work is to reveal the potentials of quantum ML and compare its performance with classical ML, the quantum approaches used are summarized in Table 1 for clarity.

**Table 1** Comparison of quantum SVR approaches implemented in this study.

| Method | Circuit Design | Kernel / Feature Strategy | Remarks |
| --- | --- | --- | --- |
| 2.2.3 Feature-based | 3 circuits | Quantum expectation values extracted as features, then fed into SVR | Generates 50–70 quantum-derived features |
| 2.2.5 Kernel-based | Minimal encoding (RY), shallow CNOT chain | Kernel = 0.3 × quantum overlap + 0.7 × classical RBF | Not a pure quantum kernel |
| 2.2.6 Fixed-Parameter | 2-layer parameterized VQC | Randomly initialized θ parameters; features extracted and used in SVR | Circuit params fixed (θ frozen), only SVR is trained |
| 2.2.7 Kernel-based | 4 circuits | Weighted ensemble of quantum state overlaps: K = 0.35K1+0.25K2+0.25K3+0.15K4 | Pure quantum kernel (no classical blending) |

## 3. RESULTS and DISCUSSION

The results indicate that the various SVR-based models show very high accuracy in $R^2$ for both training and test sets (Fig. 7). The classical SVR model achieved $R^2 = 0.938$ on the test set. While this value is acceptable for basic validation of the model, it falls short compared to the extended and quantum-based models. Specifically, the Enhanced Classical SVR and Quantum-Inspired SVR attained $R^2$ values of 0.973 and 0.971 on the test set, respectively. This shows how additional feature engineering and quantum-inspired representations improve learning performance. The Hybrid SVR and Enhanced Hybrid SVR models also performed well, showing an $R^2$ value of 0.971 on the test set. This confirms that combining classical and quantum methods provides strong generalization ability. The improved Hybrid SVR model resulted in an almost perfect fit of 0.998 on the training set, though it showed some overfitting with an $R^2$ of 0.973 on the test set. The Quantum Core SVR model performed slightly worse than the classical extended and hybrid models, achieving an $R^2$ of 0.959 on the test set. However, since the quantum kernel method offers a different measure of similarity than classical kernel functions, it proved to have strong representation ability for moderate numbers of features. The Fixed-Parameter Quantum Feature Map SVR model also performed poorly compared to the classical extended and hybrid models, with an $R^2$ of 0.955 on the test set. This might be due to the low number of qubits and the limited depth of the ansatz. However, this shows that variational quantum circuits remain sensitive to optimization and ansatz design. Finally, the Pure Quantum SVR model achieved impressive performance, scoring $R^2$ of 0.999 on the training set and 0.968 on the test set. This result highlights that pure quantum descriptions significantly help data representation, but do not differ much in test performance compared to hybrid and extended classical methods. Overall, all models recorded test $R^2$ values in the 94-99% range. The feature engineering and SVR-based modelling were quite successful in tackling a complex nuclear physics issue like fission barrier energy estimation. The main takeaway is that quantum-inspired and hybrid methods provide better generalization power than classical SVR. However, the gap between high training performance in pure quantum models and their test performance shows that quantum methods are still sensitive to optimization strategies. In this light, hybrid methods offer the most balanced solution by merging the stability of classical approaches with the advantages of quantum representations.

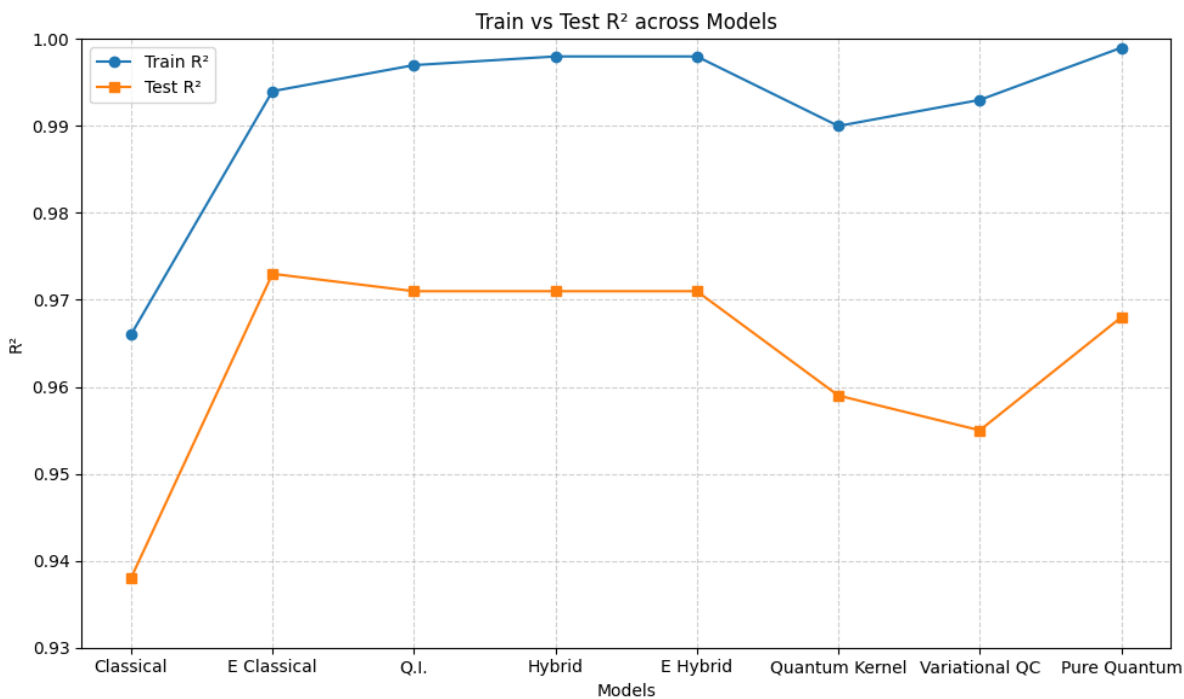

**Fig. 7** Comparison of training and testing R² values across different SVR-based models.

As shown in Fig.8 (top panel), RMSE (root mean square error) results reveal important performance differences among various SVR-based models. The classical SVR model had the highest RMSE values for both the training set (0.293) and the test set (0.366). This suggests that the small number of features used makes the model unable to effectively represent complex nuclear structures, leading to higher errors. The Enhanced Classical SVR, developed through additional feature engineering, significantly reduced the RMSE to 0.244 on the test set. This shows that the extra physics-based features boost the model's learning ability and improve the performance of classical methods. The quantum-inspired (Q-I) and hybrid SVR models performed similarly on the test set with RMSE values of 0.252 and 0.249, respectively, achieving lower error rates than the classical model. These results indicate that quantum-based feature maps help in the learning process but do not provide a significant improvement on their own. Enhanced Hybrid SVR achieved the lowest error rate with an RMSE of 0.237 on the test set, marking one of the most successful outcomes among all models. This shows that combining classical feature engineering with quantum representations offers the most balanced and effective solution. Quantum Kernel SVR (0.298) and Fixed-Parameter Quantum Feature Map SVR (0.314) had relatively higher RMSE values on the test set. This finding indicates that quantum kernel methods and variational circuits still face optimization challenges and limited parameter space. In particular, the high RMSE of Fixed-Parameter Quantum Feature Map SVR suggests that these methods struggle to represent complex structures with low qubit counts or limited ansatz depth. Finally, the Pure Quantum SVR model performed better than the classical model but slightly worse than the hybrid approaches, with an RMSE of 0.263 on the test set. This shows that pure quantum representations can reduce errors, but optimization strategies need improvement. Overall, RMSE analyses indicate that classical SVR has high errors, while hybrid and enhanced models offer significant improvements. Although quantum-based methods are promising, hybrid approaches currently seem more effective in reducing the error rate and improving generalization.

The MAE (mean absolute error) results show notable differences in the average prediction errors of the models (Fig.8 middle panel). The classical SVR model had the highest error rate, with a MAE of 0.266 on the test set. This result shows that the model cannot sufficiently reduce systematic errors because of its limited feature set. Enhanced Classical SVR significantly lowered errors compared to the classical model, achieving a MAE of 0.182 on the test set. This finding highlights the importance of including additional physics-based features to improve the model's predictive ability. Quantum-inspired SVR and Hybrid SVR had similar and significantly lower error rates, with MAEs of 0.176 and 0.180 on the test set, respectively. These results show that quantum-inspired feature maps provide a better representation than classical SVR. The hybrid approach maintains this benefit while improving general performance. Enhanced Hybrid SVR reached the lowest error value of 0.169 on the test set, showcasing the best performance among all models. This indicates that combining classical extended features with quantum-based representations offers the most stable and reliable solution. In contrast, Quantum Kernel SVR (0.239) and Fixed-Parameter Quantum Feature Map SVR (0.236) had higher MAE values on the test set. These findings indicate that quantum kernel methods and variational circuits are quite sensitive to parameter selection and optimization. Specifically, Fixed-Parameter Quantum Feature Map SVR, despite showing low error on the training set, had higher error on the test set, suggesting a tendency toward overfitting. Finally, Pure Quantum SVR performed better than classical SVR, achieving a MAE of 0.196 on the test set. However, its error rate was still higher than that of the hybrid and extended classical models. This shows the potential of pure quantum representations, but it also suggests that optimization techniques need improvement. Overall, the MAE analyses support the RMSE results in assessing error magnitude. Enhanced Hybrid SVR was the most stable and reliable model, yielding the lowest error rates for both RMSE and MAE.

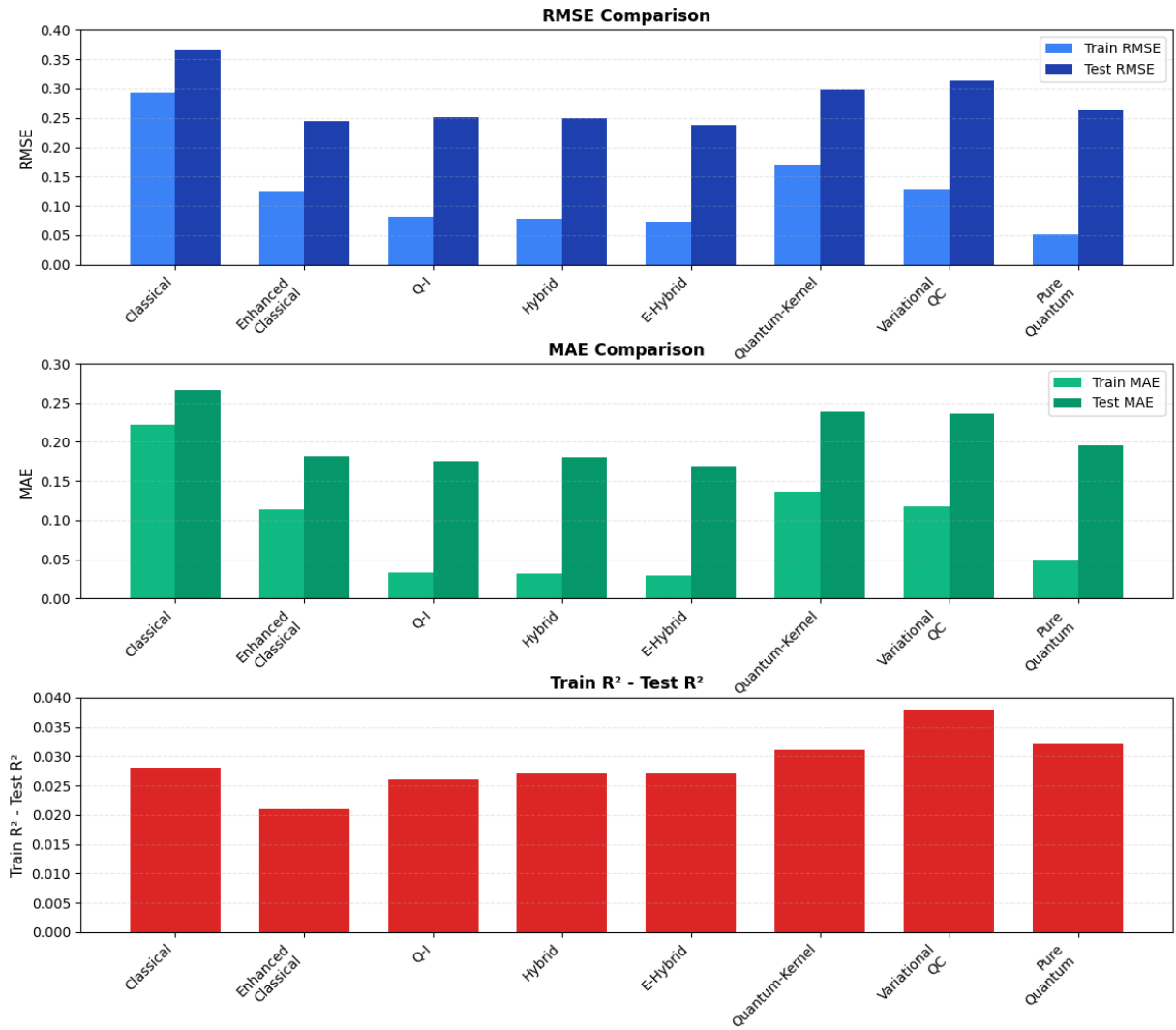

**Fig. 8** Comparative evaluation of RMSE (top), MAE (middle), and R² gap (Train–Test difference) (bottom) across SVR-based models.

The R² gap analysis, shown in the bottom panel of Fig. 8, is a key indicator for assessing how well models can generalize. Classical SVR shows a notable performance loss between training and test sets, with a difference of about 0.028. This indicates that the model has higher levels of bias and variance because it uses limited features. Enhanced Classical SVR significantly improved generalization performance by lowering the gap value to 0.021. This result shows that better feature engineering not only reduces errors but also strikes a balance between training and test performance. Quantum-inspired SVR and Hybrid SVR displayed balanced generalization performance, with gap values between roughly 0.025 and 0.026. This indicates that quantum-based representation offers more consistent results than classical SVR and reduces the risk of overfitting. Enhanced Hybrid SVR was among the models that achieved the best balance, with a gap value of 0.026. This shows that this model is one of the most reliable methods, with both low RMSE/MAE and a small gap value. In contrast, Quantum Kernel SVR (0.029) and particularly, Fixed-Parameter Quantum Feature Map SVR (0.037) stand out with their higher gap values. The high gap value in Fixed-Parameter Quantum Feature Map SVR can be attributed to the inability of the strong fit achieved in the training set to hold up in the test set. This shows the model's tendency to overfit and how sensitive it is to the parameters of variational quantum circuits. Pure Quantum SVR kept a gap value of 0.031, offering more consistent results than the classical model. However, it did not achieve the stability of the hybrid and extended models. This indicates that while pure quantum approaches show promise, they haven't yet matched the stability of hybrid strategies regarding generalization. Overall, the $R^2$ gap analysis shows that hybrid and enhanced classical models deliver the most stable generalization performance, while pure quantum and variational methods carry a greater risk of overfitting.

**CONCLUSIONS**

In this study, we compared classical, hybrid, and quantum SVR-based methods for estimating nuclear fission barrier heights. The results show that using advanced feature engineering and hybrid strategies offers the most reliable and accurate outcomes. These methods achieved the lowest RMSE and MAE values while keeping $R^2$ gaps small between training and test sets. The Improved Hybrid SVR model stood out as the most balanced and robust approach, combining the stability of classical methods with the expressive power of quantum-inspired features.

However, the analysis also points to the potential of purely quantum approaches. Pure Quantum SVR shows nearly perfect training performance and competitive test accuracy. This indicates how well quantum kernels can capture complex correlations in nuclear data. Although the Quantum Kernel SVR and Fixed-Parameter Quantum Feature Map SVR models are sensitive to circuit design and optimization issues, their results reinforce the promise of quantum methods as viable alternatives.

The tendency toward overfitting and dependence on parameters should be viewed as limitations of current quantum algorithm development rather than fundamental flaws. Overall, the findings show that quantum machine learning, especially in its hybrid and feature-rich forms, is a powerful additional tool for applications in nuclear physics. While established theoretical frameworks like FRLDM and EDF/HFB are still necessary, SVR-based quantum and hybrid methods can act as fast, general, and uncertainty-reducing predictive tools. As quantum hardware and algorithm optimization continue to progress, the ability of quantum models to predict outcomes is expected to improve and create a meaningful connection between nuclear physics theory and computational quantum technologies.

## Acknowledgements

This research was supported by the Scientific and Technological Research Council of Turkey (TUBITAK) under the BIDEB 2219-International Postdoctoral Research Scholarship Program (1059B192402323) and the UK Science and Technology Facilities Council (STFC) under grant ST/Y000358/1.